# A smart local moving algorithm for large-scale modularity-based community detection

Ludo Waltman and Nees Jan van Eck

Centre for Science and Technology Studies, Leiden University, The Netherlands
{waltmanlr, ecknjpvan}@cwts.leidenuniv.nl

We introduce a new algorithm for modularity-based community detection in large networks. The algorithm, which we refer to as a smart local moving algorithm, takes advantage of a well-known local moving heuristic that is also used by other algorithms. Compared with these other algorithms, our proposed algorithm uses the local moving heuristic in a more sophisticated way. Based on an analysis of a diverse set of networks, we show that our smart local moving algorithm identifies community structures with higher modularity values than other algorithms for large-scale modularity optimization, among which the popular 'Louvain algorithm' introduced by Blondel et al. (2008). The computational efficiency of our algorithm makes it possible to perform community detection in networks with tens of millions of nodes and hundreds of millions of edges. Our smart local moving algorithm also performs well in small and medium-sized networks. In short computing times, it identifies community structures with modularity values equally high as, or almost as high as, the highest values reported in the literature, and sometimes even higher than the highest values found in the literature.

## 1. Introduction

The problem of community detection in networks has received a lot of attention in the network science literature (Fortunato, 2010). Communities are clusters of closely connected nodes within a network. A popular approach to community detection is based on the idea of optimizing a modularity function. Modularity functions were introduced by Newman and Girvan (2004), and the idea of detecting communities by optimizing a modularity function was proposed by Newman (2004a). Nowadays, there are many variants of the modularity-based community detection approach.

These variants for instance deal with directed or weighted networks (Leicht & Newman, 2008; Newman, 2004b), or they offer a resolution parameter (Reichardt & Bornholdt, 2006) that makes it possible to customize the granularity level at which communities are detected and to mitigate the so-called resolution limit problem (Fortunato & Barthélemy, 2007). Also, some variants of the modularity-based community detection approach use modularity functions with a somewhat modified mathematical structure (e.g., Reichardt & Bornholdt, 2006; Traag, Van Dooren, & Nesterov, 2011; Waltman, Van Eck, & Noyons, 2010).

Optimizing modularity is an NP-hard problem (Brandes et al., 2008). Exact algorithms (Aloise et al., 2010; Brandes et al., 2008; Xu, Tsoka, & Papageorgiou, 2007) can be used only for small networks. Many different heuristic algorithms have been proposed for modularity optimization (Fortunato, 2010), for instance based on agglomerative hierarchical clustering (Clauset, Newman, & Moore, 2004; Newman, 2004a), simulated annealing (Guimerà, Sales-Pardo, & Amaral, 2004; Reichardt & Bornholdt, 2006), extremal optimization (Duch & Arenas, 2005), spectral optimization (Newman, 2006a, 2006b), mean field annealing (Lehmann & Hansen, 2007), and conformational space annealing (Lee, Gross, & Lee, 2012). However, most of the algorithms proposed in the literature are suitable only for small and medium-sized networks.

In this paper, we introduce an optimization algorithm that produces high-quality results even for very large networks (e.g., with tens of millions of nodes and hundreds of millions of edges). For small and medium-sized networks, our algorithm can be considered more or less competitive with the best algorithms presently available. The algorithm that we introduce builds on ideas from existing algorithms for large-scale modularity optimization (Blondel, Guillaume, Lambiotte, & Lefebvre, 2008; Rotta & Noack, 2011). These algorithms will be used as benchmarks for assessing the performance of our algorithm. We refer to our proposed algorithm as a smart local moving (SLM) algorithm. As discussed in detail below, this is because our algorithm relies heavily on a well-known local moving heuristic. Compared with existing algorithms, our SLM algorithm uses this local moving heuristic in a more sophisticated way, and it therefore produces more accurate results.

This paper is organized as follows. In Section 2, we discuss two existing algorithms for large-scale modularity optimization. We also introduce an iterative variant of both algorithms. In Section 3, we present our smart local moving algorithm.

We compare the performance of the various algorithms in Section 4. We first consider small and medium-sized networks, and we then focus on large networks. We summarize the conclusions of our research in Section 5.

## 2. Existing algorithms

Before introducing our SLM algorithm, we first discuss two existing algorithms for large-scale modularity optimization. One is the algorithm proposed by Blondel et al. (2008), often referred to as the Louvain algorithm. The other is an extension of the Louvain algorithm with a so-called multilevel refinement procedure, as proposed by Rotta and Noack (2011). Moreover, for both algorithms, we introduce an approach that allows the results of the algorithms to be improved further. Basically, this approach consists of running the algorithms in an iterative fashion, with the output of each run serving as input for the next run.

Both in this section and in the next one, we use the well-known karate club network of Zachary (1977) to illustrate the various algorithms. The karate club network represents the friendships between 34 members of a karate club.

*Modularity*

The modularity function of Newman and Girvan (2004) can be written as

$$Q = \frac{1}{2m} \sum_{i,j} \left( A_{ij} - \frac{k_i k_j}{2m} \right) \delta\left(c_i, c_j\right),$$  (1)

where $c_i$ denotes the community to which node $i$ has been assigned, $A_{ij}$ denotes whether there is an edge between nodes $i$ and $j$ ($A_{ij} = 1$) or not ($A_{ij} = 0$),

$$k_i = \sum_j A_{ij}$$  (2)

denotes the degree of node $i$, and

$$m = \frac{1}{2} \sum_{i,j} A_{ij}$$  (3)

denotes the total number of edges in the network. The function $\delta(c_i, c_j)$ indicates whether nodes $i$ and $j$ belong to the same community. It equals 1 if $c_i = c_j$ and 0 otherwise. The modularity function in (1) also applies to weighted networks (Newman, 2004b). The only difference is that in the case of a weighted network $A_{ij}$ may take any non-negative value. Higher values of the modularity function in (1) are supposed to indicate a better community structure. Given a network of $n$ nodes, the idea of modularity-based community detection therefore is to try to find values of $c_1$, ..., $c_n$ that maximize (1). These values of $c_1$, ..., $c_n$ are considered to represent the optimal community structure for the given network.

A number of variants of the modularity function in (1) have been proposed in the literature. These variants for instance include a resolution parameter (Reichardt & Bornholdt, 2006), or they have a somewhat modified mathematical structure (e.g., Reichardt & Bornholdt, 2006; Traag et al., 2011; Waltman et al., 2010). In general, optimization techniques developed for the original modularity function can also be applied to the alternative modularity functions discussed in the literature. In this paper, our focus is on the optimization of the original modularity function, but our proposed approach extends to other modularity functions as well.

*Local moving heuristic*

A frequently used approach to modularity optimization is the local moving heuristic. The idea of the local moving heuristic is to repeatedly move individual nodes from one community to another in such a way that each node movement results in a modularity increase. The local moving heuristic iterates over the nodes in a network in a random order. For each node, it is determined whether it is possible to increase modularity by moving the node from its current community to a different (possibly empty) community. If increasing modularity is indeed possible, the node is moved to the community that results in the largest modularity gain. The local moving heuristic keeps moving nodes until a situation is reached in which there are no further possibilities to increase modularity through individual node movements. The local moving heuristic has been quite popular in the literature (Barber & Clark, 2009; Blondel et al., 2008; Liu & Murata, 2010; Mei, He, Shi, Wang, & Li, 2009; Rotta & Noack, 2011; Schuetz & Caflisch, 2008; Ye, Hu, & Yu, 2008), probably in part because it can be implemented in an efficient way (Blondel et al., 2008). The local moving heuristic plays a central role both in the two existing algorithms for large-scale modularity optimization discussed below and in our SLM algorithm introduced in the next section.

Figure 1 shows the karate club network, using colors to indicate an example of a community structure that has been obtained using the local moving heuristic. Starting from a situation in which each of the 34 nodes in the network belongs to its own community, the local moving heuristic has identified a solution in which the nodes are organized into six communities. This solution has a modularity value of 0.3791. The solution has the property that it is not possible to increase modularity by moving an individual node from one community to another. In other words, the solution is locally optimal with respect to individual node movements. We emphasize that the solution shown in Figure 1 is not unique. Depending on the order in which the local moving heuristic iterates over the nodes in the network, other solutions may be obtained as well.

*Louvain algorithm*

The Louvain algorithm proposed by Blondel et al. (2008) starts with each node in a network belonging to its own community. So initially each community is a singleton, consisting of one node only. The algorithm then uses the local moving heuristic to obtain an improved community structure. Hence, individual nodes are moved from one community to another until no further increase in modularity can be achieved. At this point, a reduced network is constructed (Arenas, Duch, Fernández, & Gómez, 2007). This is a network in which each node corresponds with a community in the original network. In the reduced network, the weight of an edge between two nodes equals the total weight of all edges between the nodes in the two corresponding communities in the original network. Edges between nodes in the same community in the original network result in self links in the reduced network. As a consequence of the close relation between the original network and the reduced one, merging communities in the original network is equivalent to grouping the corresponding nodes in the reduced network together in a community.

The Louvain algorithm proceeds by assigning each node in the reduced network to its own singleton community. Next, the local moving heuristic is applied in the reduced network, in exactly the same way as was done before in the original network. Based on the resulting community structure, a second reduced network is constructed. This network is treated in the same way as the original network and the first reduced network. Hence, the local moving heuristic is applied and another reduced network is constructed. The Louvain algorithm continues in this way until a network is obtained that cannot be reduced further. One now has a sequence of successively smaller networks. This sequence of networks corresponds with a sequence of mergers of smaller communities into larger ones, and in this way it determines the final assignment of the nodes in the original network to communities.

Figure 2 summarizes the main steps of the Louvain algorithm.[1] As can be seen, the algorithm can be conveniently written in a recursive form. The initial assignment of nodes to communities is not specified in Figure 2. However, as explained above, the Louvain algorithm normally starts with each node in a network belonging to its own community. We will get back to this below.

Figure 3 illustrates the application of the Louvain algorithm to the karate club network. First, the local moving heuristic is used. Suppose this gives the community structure shown in Figure 1. As already mentioned, this community structure has a modularity value of 0.3791. Figure 3(a) shows the reduced network corresponding with the community structure shown in Figure 1. The result of applying the local moving heuristic in the reduced network is shown in Figure 3(b). As can be seen, the local moving heuristic has assigned nodes B and C in the reduced network to the same community. The same holds for nodes E and F. The community structure shown in Figure 3(b) has a modularity value of 0.4151. Based on this community structure, a second reduced network can be constructed and again the local moving heuristic can be applied. However, it turns out that no further increase in modularity is possible. Figure 3(c) shows the final community structure in the original network.

*Louvain algorithm with multilevel refinement*

When a community structure has been obtained using the Louvain algorithm, one can be sure that the community structure cannot be improved further by merging communities. In other words, the Louvain algorithm finds solutions that are locally optimal with respect to community merging. However, solutions found by the Louvain algorithm need not be locally optimal with respect to movements of individual nodes between communities. It may be possible to improve a solution by moving an individual node from one community to another.

---

[1] In the pseudocode presented in this paper, we use a MATLAB-like syntax when dealing with arrays. For instance, $x(y = i) \leftarrow j$, where $x$ and $y$ are arrays of equal length and $i$ and $j$ are scalars, indicates that the elements of $x$ for which the corresponding elements of $y$ are equal to $i$ are selected and are set equal to $j$.

An extension of the Louvain algorithm with a multilevel refinement procedure was proposed by Rotta and Noack (2011). The multilevel refinement procedure improves solutions found by the Louvain algorithm in such a way that they become locally optimal with respect to individual node movements. To accomplish this, the local moving heuristic is used not only for creating an initial community structure for the nodes in a network but also for refining the final community structure. Moreover, this is done not only at the level of the original network but also at the level of each of the reduced networks.

A summary of the main steps of the Louvain algorithm with multilevel refinement is provided in Figure 4. As can be seen, the Louvain algorithm with multilevel refinement is identical to the original Louvain algorithm except that at each level of the recursion the local moving heuristic is used twice instead of once. Like in the original Louvain algorithm, the local moving heuristic is used for creating an initial community structure, but in addition it is also used for refining the final community structure. In this way, it is guaranteed that a community structure is obtained that cannot be improved further by moving individual nodes from one community to another.

In the case of our karate club example, extending the Louvain algorithm with the multilevel refinement procedure has the effect that, after the result shown in Figure 3(c) has been obtained, the local moving heuristic is applied a second time in the original network. It turns out that modularity can be increased by moving node 28 from the green to the blue community. Next, another modularity increase is possible by also moving node 24 to the blue community. This yields the community structure shown in Figure 5. This community structure corresponds with a modularity value of 0.4198. No further increase in modularity turns out to be possible through individual node movements. In fact, the community structure shown in Figure 5 is known to be globally optimal (Aloise et al., 2010).

*Iterative variant of the Louvain algorithm*

So far, we have discussed two algorithms for large-scale modularity optimization: The original Louvain algorithm of Blondel et al. (2008) and the Louvain algorithm extended with the multilevel refinement procedure of Rotta and Noack (2011). We now introduce an approach that aims to further improve the results of both algorithms.

The basic idea of our proposed approach is to run the algorithms in an iterative way, where the output of each iteration is used as input for the next iteration. In the case of the original Louvain algorithm, we start by assigning each node in a network to its own community, and we then run the algorithm as specified in Figure 2. This is the first iteration of the algorithm. After the first iteration has been completed, we run the algorithm a second time, but this time we do not start with each node belonging to its own community. Instead, we start with the community structure obtained in the first iteration of the algorithm. This means that the second iteration allows individual nodes to move between the communities produced in the first iteration. After the second iteration has been completed, we run the algorithm a third time, using the community structure obtained in the second iteration as input. In this way, more and more iterations of the algorithm can be performed. The iterative approach can be stopped as soon as performing an additional iteration of the algorithm does not result in a modularity increase. Alternatively, the approach can be stopped after a certain maximum number of iterations. Of course, the same iterative approach can also be applied to the Louvain algorithm with multilevel refinement.

As we have discussed, the Louvain algorithm finds solutions that are locally optimal with respect to community merging, but these solutions need not be locally optimal with respect to individual node movements. On the other hand, solutions found by the Louvain algorithm with multilevel refinement are locally optimal with respect to individual node movements, but they need not be locally optimal with respect to community merging. However, when our iterative approach is applied to the Louvain algorithm, either with or without multilevel refinement, it becomes possible to find solutions that are locally optimal with respect to both community merging and individual node movements. When the iterative approach has converged (i.e., the last iteration did not result in a modularity increase), one can be sure to have a community structure that cannot be improved further either by merging communities or by moving individual nodes from one community to another.

## 3. Smart local moving algorithm

Using the above discussed iterative approach, the Louvain algorithm, either with or without multilevel refinement, identifies solutions that are locally optimal with respect to both community merging and individual node movements. Solutions will in general not be locally optimal with respect to community splitting or with respect to movements of sets of nodes from one community to another. Like the iterative variant of the Louvain algorithm, the SLM algorithm that we introduce in this section identifies solutions that are locally optimal with respect to both community merging and individual node movements. In addition, however, the SLM algorithm also searches for possibilities to increase modularity by splitting up communities and by moving sets of nodes from one community to another. As we will see, this is accomplished by using the local moving heuristic in a more sophisticated way than is done in the Louvain algorithm.

Like the Louvain algorithm, the SLM algorithm starts with each node in a network being assigned to its own singleton community. Also like the Louvain algorithm, the SLM algorithm uses the local moving heuristic to obtain an improved community structure. However, after the local moving heuristic has been applied, the SLM algorithm takes a different approach than the Louvain algorithm. As we have seen in the previous section, the Louvain algorithm proceeds by constructing a reduced network. The SLM algorithm will also construct a reduced network, but before it does so, it first takes some other steps.

The SLM algorithm iterates over all communities in the present community structure. For each community, a so-called subnetwork is constructed. This is a copy of the original network that includes only the nodes belonging to the specific community of interest. The SLM algorithm then uses the local moving heuristic to identify communities in the subnetwork. Each node in the subnetwork is first assigned to its own singleton community, and then the local moving heuristic is applied. In some cases, this yields a community structure consisting of one big community that includes all nodes in the subnetwork. In other cases, a community structure is obtained consisting of multiple communities that each include some of the nodes in the subnetwork.

After a community structure has been obtained for each of the subnetworks, the SLM algorithm constructs a reduced network. In the reduced network, each node corresponds with a community in one of the subnetworks. The SLM algorithm then performs an initial assignment of the nodes in the reduced network to communities. Nodes corresponding with communities in the same subnetwork are assigned to the same community in the reduced network. Hence, for each subnetwork, there is one community in the reduced network.[2]

At this point, the entire process starts all over again, but this time based on the reduced network rather than the original one. So first the local moving heuristic is applied in the reduced network, and then for each community in the reduced network a subnetwork is constructed and communities in the subnetwork are identified. This is the starting point for the construction of a second reduced network, after which the entire process again repeats itself. The SLM algorithm moves on in this way until a network is obtained that cannot be reduced further.

Figure 6 offers a summary of the main steps of the SLM algorithm. The algorithm is again written in a recursive form. The SLM algorithm is similar to the Louvain algorithm outlined in Figure 2 except that, due to the idea of applying the local moving heuristic at the level of subnetworks, communities can be split up and sets of nodes can be moved from one community to another. In this way, the SLM algorithm has more freedom in searching for high-quality solutions to the modularity optimization problem.

To illustrate the SLM algorithm, we again consider our karate club example. We first go back to Figure 1. This figure shows the community structure obtained by applying the local moving heuristic in the original network. There are six communities, each indicated using a different color. As explained above, after applying the local moving heuristic in the original network, for each community a subnetwork is constructed. The six subnetworks that are obtained in this way are shown in Figure 7(a). The local moving heuristic is applied in each of these subnetworks. In the green, blue, purple, and yellow subnetworks, this results in all nodes being assigned to the same community. This is not the case for the red and orange subnetworks. These subnetworks are each split up into two communities. In Figure 7(a), this is indicated by displaying some nodes using circles and others using squares.

---

[2] Notice that the nodes in the reduced network are not assigned to singleton communities. This would result in a decrease in modularity in the case in which a subnetwork has a community structure consisting of multiple communities while in fact having one big community would be better from a modularity point of view. Assigning nodes corresponding with communities in the same subnetwork to the same community in the reduced network guarantees that modularity increases monotonically in the SLM algorithm.

Figure 7(b) shows the reduced network that is obtained. There are eight nodes in the reduced network, one for each community in a subnetwork. Nodes corresponding with communities in the same subnetwork are initially assigned to the same community in the reduced network. The result of applying the local moving heuristic in the reduced network is shown in Figure 7(c). As can be seen, nodes A1 and A2 in the reduced network have remained in the same community. However, node C1, which initially was in a community with node C2, has been assigned to the same community as node D. The next step is to construct subnetworks based on the community structure shown in Figure 7(c), to apply the local moving heuristic in each subnetwork, to construct a second reduced network, and to apply the local moving heuristic in this network. However, we do not show any further results, since it turns out that the community structure shown in Figure 7(c) cannot be improved further. The corresponding community structure in the original network is shown in Figure 5. Hence, in this particular example, the SLM algorithm identifies the same community structure as the Louvain algorithm with multilevel refinement. As mentioned before, this community structure is known to be globally optimal.

Like the Louvain algorithm, the SLM algorithm can be run in an iterative way. In the first iteration of the algorithm, we start from an initial situation in which each node in a network is assigned to its own community. In the second iteration, we start with nodes being assigned to the communities obtained in the first iteration. In the third iteration, the communities obtained in the second iteration are our starting point, and so on. There is one important difference with the iterative variant of the Louvain algorithm. At some point, the iterative variant of the Louvain algorithm, either with or without multilevel refinement, converges. This happens when a community structure is obtained that cannot be improved further either by merging communities or by moving individual nodes from one community to another. In the case of the iterative variant of the SLM algorithm, there is no convergence like this. When running the SLM algorithm in an iterative way, the algorithm keeps searching for possibilities to increase modularity by splitting up communities and by moving sets of nodes from one community to another. Hence, in the case of the iterative variant of the SLM algorithm, it may always be possible to obtain further improvements in community structure by performing more iterations of the algorithm.

## 4. Results

In this section, we compare the performance of our SLM algorithm with the performance of the Louvain algorithm, both with and without multilevel refinement. The performance of the algorithms is compared using 13 small and medium-sized networks and six large networks. In the case of the small and medium-sized networks, we also make a comparison with the best results reported in the literature. All networks that we use are unweighted and undirected and do not have loops. Some networks have more than one connected component. In that case, all connected components are included, not only the largest one.

The results presented in this section were obtained using our own implementations of the original Louvain algorithm, the Louvain algorithm with multilevel refinement, and the SLM algorithm. The algorithms were implemented in Java. The implementations along with some documentation can be downloaded from www.ludowaltman.nl/slm/. Although in this paper our focus is on unweighted networks, the implementations also support weighted networks. In addition, support is offered for a resolution parameter (Reichardt & Bornholdt, 2006) that can be used to customize the granularity level at which communities are detected.

All calculations reported below were performed on a system with an Intel Xeon CPU (L5520 @ 2.27 GHz) and 64 GB internal memory.

*Results for small and medium-sized networks*

To analyze the performance of our SLM algorithm in the case of small and medium-sized networks, we take an approach that is similar to the one used by Lee at al. (2012). Lee et al. introduced a new algorithm for modularity optimization referred to as conformational space annealing (CSA). They used 13 small and medium-sized networks to evaluate the performance of their algorithm. It turned out that for each of the 13 networks the CSA algorithm was able to identify a solution with a modularity value higher than or equal to the highest modularity value reported in the literature.

Because of the excellent performance of the CSA algorithm, we use it as a benchmark for assessing the performance of the SLM algorithm. We also use the same 13 networks as were used by Lee et al. (2012). The first columns of Table 1 report for each of these networks the number of nodes and edges as well as the highest modularity value obtained using the CSA algorithm. The networks are listed in increasing order of their number of nodes. The modularity values reported in Table 1 for the CSA algorithm have been taken from Tables 2 and 3 in the paper by Lee et al.

For each of the 13 networks, we tested three algorithms: The original Louvain algorithm, the Louvain algorithm with multilevel refinement, and the SLM algorithm. Each algorithm was run 100 times using different random numbers.[3] Each algorithm run consisted of 100 iterations. For each combination of a network and an algorithm, we determined the highest modularity value obtained at the end of the 100th iteration of the 100 algorithm runs. In addition, in order to analyze the effect of performing multiple iterations of an algorithm, we also determined the highest modularity value obtained at the end of the first iteration of the 100 algorithm runs.

All results are reported in Table 1. The table does not show the modularity values themselves. Instead, the table shows the difference between a modularity value and the corresponding modularity value obtained using the CSA algorithm. Negative values indicate that an algorithm is performing worse than the CSA algorithm. Positive values indicate a better performance than the CSA algorithm. In the case of an equal performance, no value is shown.

Based on Table 1, the following observations can be made:

- In the case of the smallest networks, the original Louvain algorithm, the Louvain algorithm with multilevel refinement, and the SLM algorithm all perform equally well as the CSA algorithm, even with only one iteration per algorithm run. In fact, in the case of the first four networks listed in Table 1, all algorithms identify solutions with modularity values equal to those obtained using exact algorithms (Aloise et al., 2010).

- With 100 iterations per algorithm run, the SLM algorithm can be considered more or less competitive with the CSA algorithm. There are four networks for which the SLM algorithm performs worse than the CSA algorithm, but the differences are small (at most 0.2% difference in modularity value). On the other hand, there is one network for which the SLM algorithm outperforms the CSA algorithm, but again the difference is not very large (0.5% difference in modularity value). Except for the smallest networks discussed above, the SLM algorithm consistently outperforms the original Louvain algorithm and the Louvain algorithm with multilevel refinement, with differences in modularity value of at most 1.0%. Notice that the original Louvain algorithm and the Louvain algorithm with multilevel refinement have (almost) the same performance.

---

[3] These random numbers determine the order in which the local moving heuristic iterates over the nodes in a network. Different random numbers result in a different order in which the nodes in a network are visited. Depending on the order in which the nodes in a network are visited, the local moving heuristic may identify different community structures.

- With only one iteration per algorithm run, the SLM algorithm performs significantly worse than the CSA algorithm (except for the smallest networks). Moreover, the SLM algorithm is also significantly outperformed by the Louvain algorithm with multilevel refinement. The SLM algorithm performs at about the same level as the original Louvain algorithm. Notice that the Louvain algorithm with multilevel refinement has (almost) the same performance regardless of the number of iterations (1 or 100) per algorithm run.

In summary, it can be concluded that in the case of small and medium-sized networks the SLM algorithm is able to compete with the best algorithms presently available, but in order to do so it is crucial to perform a sufficiently large number of iterations per algorithm run.

The importance of performing a sufficiently large number of iterations of the SLM algorithm is also illustrated in Figure 8. Based on 1300 runs of the SLM algorithm (i.e., 100 runs for each of the 13 networks), the figure shows for each iteration the percentage of all runs that resulted in a modularity increase. The same statistics are also reported for the original Louvain algorithm and for the Louvain algorithm with multilevel refinement. In the case of the original Louvain algorithm, it turns out that after four iterations all 1300 algorithm runs had converged. In the case of the Louvain algorithm with multilevel refinement, it took only three iterations for all 1300 algorithm runs to converge. The results obtained using the SLM algorithm are quite different. As discussed in Section 3, the SLM algorithm keeps searching for possibilities to increase modularity. Indeed, Figure 8 shows that in iteration 10 still about 19% of the 1300 runs of the SLM algorithm resulted in a modularity increase. Even in iteration 100, a modularity increase still took place in almost 2% of the algorithm runs.

Finally, let us consider the issue of computing time. It turns out that in terms of computing time the SLM algorithm compares quite favorably with the CSA algorithm. The total time required to perform 100 runs of the SLM algorithm was less than 10 seconds for the nine smallest networks (in terms of number of nodes), less than one minute for the E-mail and Erdos02 networks, and less than two minutes for the PGP network. For the condmat2003 network, the largest network among our 13 small and medium-sized networks, it took 555 seconds to perform 100 runs of the SLM algorithm. Because of the use of different computer systems, these computing times are not directly comparable with the ones reported by Lee et al. (2012, Table 2) for the CSA algorithm. Nevertheless, it is clear that for larger networks the SLM algorithm is computationally much more efficient than the CSA algorithm. In the case of the condmat2003 network, for instance, 50 runs of the CSA algorithm require about 100 times more computing time than 100 runs of the SLM algorithm (57 609 vs. 555 seconds).

From a computational point of view, the original Louvain algorithm and the Louvain algorithm with multilevel refinement perform even better than the SLM algorithm, especially when working with somewhat larger networks. For instance, in the case of the condmat2003 network, these algorithms require only about 25% of the computing time of the SLM algorithm. Of course, the difference in computing time between the SLM algorithm and the Louvain algorithm strongly depends on the number of iterations performed per algorithm run, since the Louvain algorithm tends to converge after a few iterations while the SLM algorithm keeps trying to find possibilities to increase modularity. Below, in our analysis of large networks, we will compare the computational performance of the SLM algorithm, the original Louvain algorithm, and the Louvain algorithm with multilevel refinement in more detail.

*Results for large networks*

The main focus of our SLM algorithm is on community detection in large and very large networks. We have selected six large networks, originating from a number of different domains, to analyze the large-scale performance of the SLM algorithm. The following networks are considered:

- *Amazon*. Network of frequently co-purchased products on the Amazon website (Yang & Leskovec, 2012).

- *DBLP*. Co-authorship network obtained from the DBLP computer science bibliography (Yang & Leskovec, 2012).

- *IMDb*. Network of actors playing in the same movie obtained from the Internet Movie Database (Barabási & Albert, 1999).

- *LiveJournal*. Friendship network of the LiveJournal online blogging community (Yang & Leskovec, 2012).

- *WoS*. Citation network of all scientific articles in the Web of Science database in the period 2002–2011. This network is similar to the citation network that we studied in an earlier paper (Waltman & Van Eck, 2012).

- *Web uk-2005*. Web network obtained from a crawl of the .uk domain in 2005. The crawl was performed using UbiCrawler (Boldi, Codenotti, Santini, & Vigna, 2004), and the network is made available by the Laboratory for Web Algorithmics at http://law.di.unimi.it. The network was also used by Blondel et al. (2008) to evaluate the performance of the Louvain algorithm.

Table 2 shows the number of nodes and edges in each of the above networks. The number of nodes ranges between 0.4 million (DBLP and IMDb) and 39.5 million (Web uk-2005). The number of edges ranges between 0.9 million (Amazon) and 783.0 million (Web uk-2005).

Like in the case of the small and medium-sized networks, we compare the SLM algorithm with the original Louvain algorithm and with the Louvain algorithm with multilevel refinement. Since community detection in large networks can be computationally quite expensive, the number of algorithm runs that were performed is smaller than in the case of the small and medium-sized networks. Instead of 100 runs, for each of the six large networks we performed 10 runs of each algorithm. Moreover, each algorithm run consisted of 10 rather than 100 iterations. Modularity values were calculated at the end of the first and the 10th iteration of each algorithm run. For each combination of a network and an algorithm, we report not only the highest modularity value obtained in 10 algorithm runs but also the lowest one. In the case of large networks, it may in practice not always be feasible to perform multiple runs of an algorithm. The lowest modularity value obtained in 10 runs of an algorithm provides an indication of the worst-case performance that can be expected when the algorithm is run only once.

The modularity values obtained for the six large networks are reported in Table 2. Computing times are reported as well. For each combination of a network and an algorithm, the table shows the average number of seconds it took to perform one algorithm run.

Our observations based on Table 2 can be summarized as follows:

- With 10 iterations per algorithm run, the SLM algorithm consistently outperforms the original Louvain algorithm and the Louvain algorithm with multilevel refinement. The difference in modularity value is largest for the DBLP network (more than 1%) and almost negligible for the Web uk-2005 network (about 0.04%). Interestingly, for all networks except IMDb, the worst run of the SLM algorithm still gives better results than the best run of each of the other two algorithms. Like in the case of the small and medium-sized networks, the original Louvain algorithm and the Louvain algorithm with multilevel refinement have (almost) the same performance.

- With only one iteration per algorithm run, the SLM algorithm slightly outperforms the original Louvain algorithm, but the difference is almost negligible. On the other hand, the SLM algorithm generally performs worse than the Louvain algorithm with multilevel refinement, sometimes with a quite significant modularity difference of more than 1%. Notice that the performance of the Louvain algorithm with multilevel refinement is hardly affected by the number of iterations (1 or 10) per algorithm run.

- In terms of computing time, when only one iteration per algorithm run is performed, the SLM algorithm is about equally expensive as the original Louvain algorithm and in general somewhat less expensive than the Louvain algorithm with multilevel refinement. When performing 10 iterations per algorithm run, the original Louvain algorithm and the Louvain algorithm with multilevel refinement require more or less the same amount of computing time and the SLM algorithm requires considerably more. In the case of the IMDb network, the SLM algorithm even needs almost four times as much computing time as the other two algorithms. The relative difference in computing time is smallest for the Web uk-2005 network, which is the largest network in our analysis. In the case of this network, a run of the SLM algorithm on average takes almost five hours, which is about 45% more than a run of the original

Louvain algorithm and about 70% more than a run of the Louvain algorithm with multilevel refinement.

Based on the above observations, it is clear that the SLM algorithm is able to identify better community structures, in terms of modularity, than the original Louvain algorithm and the Louvain algorithm with multilevel refinement. To identify high-quality community structures, it is essential to use the iterative variant of the SLM algorithm. This is in line with our findings for small and medium-sized networks. From the point of view of computing time, the iterative variant of the SLM algorithm is more expensive than the iterative variants of the other two algorithms. However, it turns out that a single run of the SLM algorithm typically gives better results than multiple runs of the other two algorithms, meaning that in the case of the SLM algorithm there is less need to perform multiple algorithm runs. Hence, although from a computational perspective a single run of the SLM algorithm is relatively expensive, this is counterbalanced by the fact that fewer algorithm runs need to be performed.

Figure 9 offers some additional insight into the effect of performing multiple runs and multiple iterations of the SLM algorithm. For two networks, DBLP and WoS, the figure shows the modularity value at the end of each of the 10 iterations in each of the 10 runs of the SLM algorithm. For both networks, the figure also shows the highest modularity value obtained using the iterative variants of the original Louvain algorithm and the Louvain algorithm with multilevel refinement. We note that in the case of the latter two algorithms convergence always took place within at most four iterations.

As can be seen in Figure 9, in the case of the DBLP and WoS networks, modularity increases in each iteration of the SLM algorithm, but after the second iteration increases in modularity tend to be relatively small. For both networks, the highest modularity value at the end of the second iteration turns out to be higher than the highest modularity value obtained using the Louvain algorithm. (Although not shown in Figure 9, the same observation can be made for the other four large networks included in our analysis.) In the case of the DBLP network, even the lowest modularity value at the end of the second iteration is higher than the highest modularity value obtained using the Louvain algorithm. (The same observation can be made for the Amazon and Web uk-2005 networks.) In the case of the WoS network, it takes five iterations before the lowest modularity value obtained using the SLM algorithm exceeds the highest modularity value obtained using the Louvain algorithm. The general picture emerging from Figure 9 is that a few iterations of the SLM algorithm are usually sufficient to outperform the Louvain algorithm. Additional iterations of the SLM algorithm lead to further increases in modularity, but the gain tends to be relatively small.

## 5. Conclusions

In this paper, we have introduced our SLM algorithm for modularity-based community detection. Our algorithm is intended primarily for community detection in large networks, and we have therefore focused on comparing our algorithm with two other algorithms for large-scale modularity-based community detection: The Louvain algorithm proposed by Blondel et al. (2008) and an extension of the Louvain algorithm with a so-called multilevel refinement procedure, as suggested by Rotta and Noack (2011). In addition to introducing a new algorithm, we have also proposed iterative variants of the original Louvain algorithm and of the Louvain algorithm with multilevel refinement.

Despite being interested mostly in community detection in large networks, we have also analyzed the performance of our SLM algorithm in small and medium-sized networks. In the case of these networks, we have compared the SLM algorithm not only with the original Louvain algorithm and the Louvain algorithm with multilevel refinement but also with the CSA algorithm introduced by Lee et al. (2012). The CSA algorithm is among the best algorithms presently available for modularity optimization in small and medium-sized networks. For a range of different networks, Lee et al. show that the CSA algorithm is able to identify community structures with modularity values higher than or equal to the highest values reported in the literature. Because of its computational demands, the CSA algorithm does not seem suitable for modularity optimization in large networks.

Based on an analysis involving 13 small and medium-sized networks and six large and very large networks (with up to 40 million nodes and up to 800 million edges), we conclude that our SLM algorithm consistently outperforms the original Louvain algorithm and the Louvain algorithm with multilevel refinement. Only in the case of very small networks, we find that all three algorithms perform equally well. The excellent results of the SLM algorithm are obtained only if in each run of the algorithm a sufficiently large number of iterations are performed. Compared with the original Louvain algorithm and the Louvain algorithm with multilevel refinement, the SLM algorithm then turns out to require considerably more computing time to perform a single algorithm run. However, this is counterbalanced by the fact that in the case of the SLM algorithm there is less need to perform multiple algorithm runs. In the analysis of the six large networks, we find that a single run of the SLM algorithm almost always yields a higher modularity value than 10 runs of the original Louvain algorithm or the Louvain algorithm with multilevel refinement.

In the case of the 13 small and medium-sized networks, we find that our SLM algorithm can be considered more or less competitive with the CSA algorithm of Lee et al. (2012). There are four networks for which the CSA algorithm gives slightly better results than the SLM algorithm, but there is also one network for which the SLM algorithm yields better results. Furthermore, for the medium-sized networks, the SLM algorithm turns out to require much less computing time than the CSA algorithm. This means that there may be room to further improve the performance of the SLM algorithm by performing more algorithms run and more iterations per algorithm run.

Finally, let us note that in this paper we have restricted ourselves to the use of the SLM algorithm for optimizing the original modularity function of Newman and Girvan (2004). We emphasize, however, that the SLM algorithm can also be used for optimizing many of the variants of this function that have been proposed in the literature, for instance variants that include a resolution parameter (Reichardt & Bornholdt, 2006) or that have a somewhat modified mathematical structure (e.g., Reichardt & Bornholdt, 2006; Traag et al., 2011; Waltman et al., 2010).

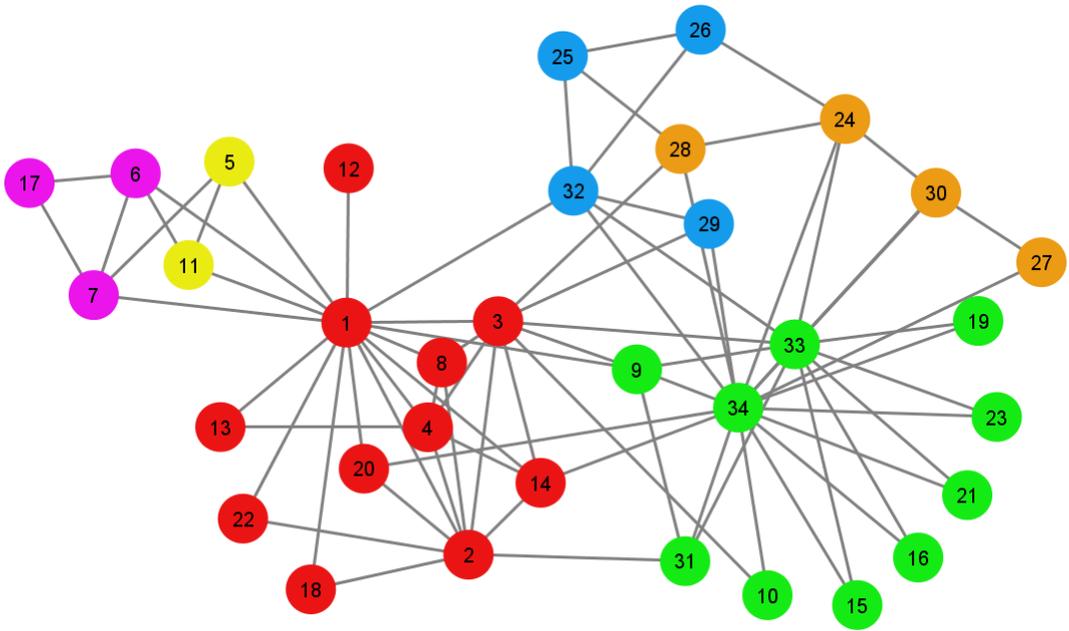

Figure 1. Result of applying the local moving heuristic to the karate club network.

**LouvainAlgorithm**

**input:**
    *A*: Adjacency matrix of a network
    *c*: Initial assignment of nodes to communities
**output:**
    *c*: Final assignment of nodes to communities

*// Run the local moving heuristic.*
$c \leftarrow$ LocalMovingHeuristic($A$, $c$)

**if** NumberOfCommunities($c$) < NumberOfNodes($A$) **then**
    *// Construct a reduced network.*
    $A_{reduced} \leftarrow$ ReducedNetwork($A$, $c$)
    $c_{reduced} \leftarrow [1 \ldots$ NumberOfNodes($A_{reduced}$)]

    *// Perform a recursive call to identify the community structure of the reduced network.*
    $c_{reduced} \leftarrow$ LouvainAlgorithm($A_{reduced}$, $c_{reduced}$)

    *// Merge communities based on the community structure of the reduced network.*
    $c_{old} \leftarrow c$
    **for** $i \leftarrow 1$ **to** NumberOfCommunities($c_{old}$) **do**
        $c(c_{old} = i) \leftarrow c_{reduced}(i)$
    **end for**
**end if**

Figure 2. Summary of the main steps of the Louvain algorithm.

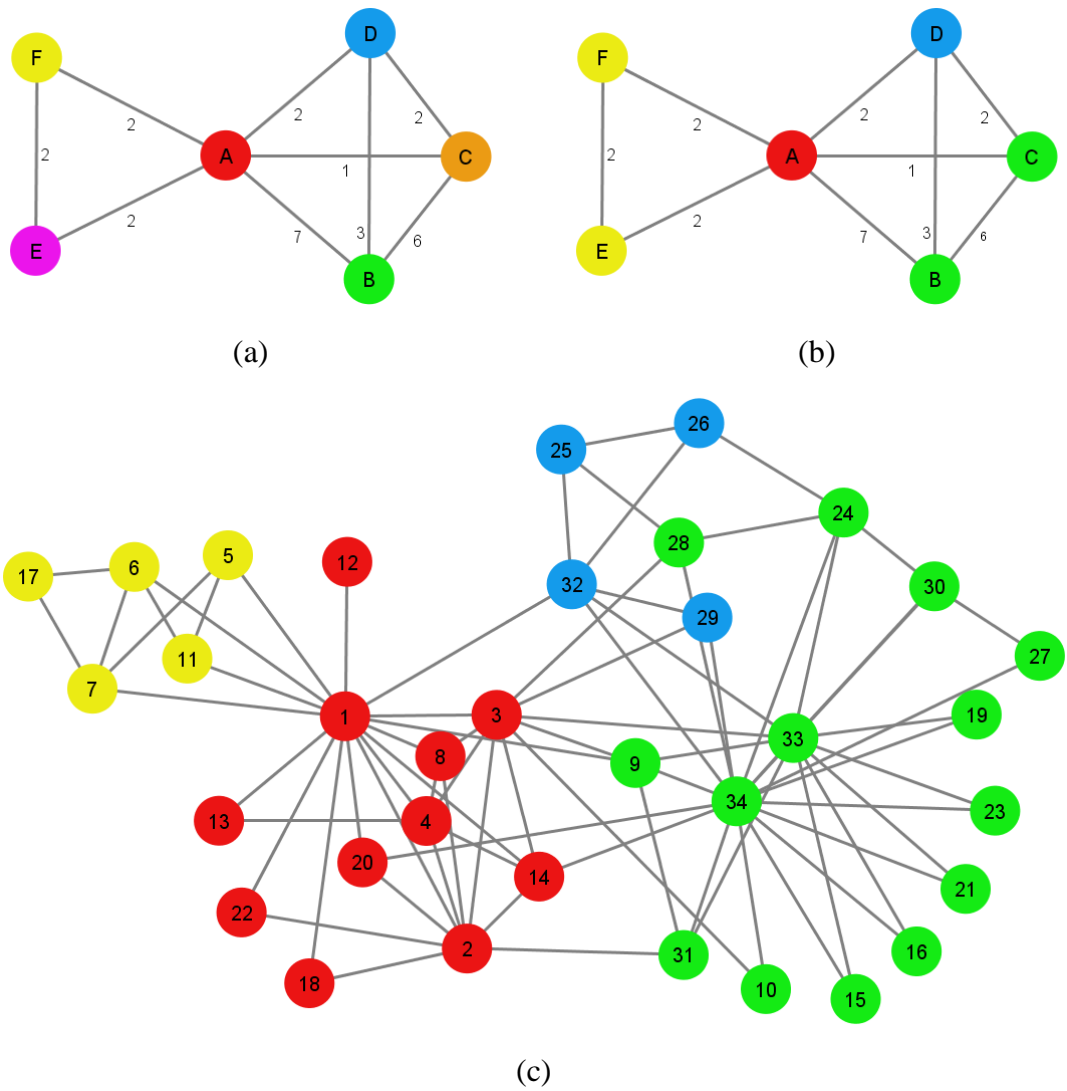

(a)                                                                              (b)

(c)

Figure 3. Result of applying the Louvain algorithm to the karate club network. (a) Reduced network before applying the local moving heuristic. (b) Reduced network after applying the local moving heuristic. (c) Final solution in the original network. Notice that self links in the reduced network are not shown.

**LouvainAlgorithmWithMultilevelRefinement**

**input:**

    *A*: Adjacency matrix of a network

    *c*: Initial assignment of nodes to communities

**output:**

    *c*: Final assignment of nodes to communities

*// Run the local moving heuristic.*

$c \leftarrow$ LocalMovingHeuristic($A$, $c$)

**if** NumberOfCommunities($c$) < NumberOfNodes($A$) **then**

    *// Construct a reduced network.*

    $A_{reduced} \leftarrow$ ReducedNetwork($A$, $c$)

    $c_{reduced} \leftarrow [1\ldots$NumberOfNodes($A_{reduced}$)]

    *// Perform a recursive call to identify the community structure of the reduced network.*

    $c_{reduced} \leftarrow$ LouvainAlgorithmWithMultilevelRefinement($A_{reduced}$, $c_{reduced}$)

    *// Merge communities based on the community structure of the reduced network.*

    $c_{old} \leftarrow c$

    **for** $i \leftarrow 1$ **to** NumberOfCommunities($c_{old}$) **do**

        $c(c_{old} = i) \leftarrow c_{reduced}(i)$

    **end for**

    *// Run the local moving heuristic.*

    $c \leftarrow$ LocalMovingHeuristic($A$, $c$)

**end if**

Figure 4. Summary of the main steps of the Louvain algorithm with multilevel refinement.

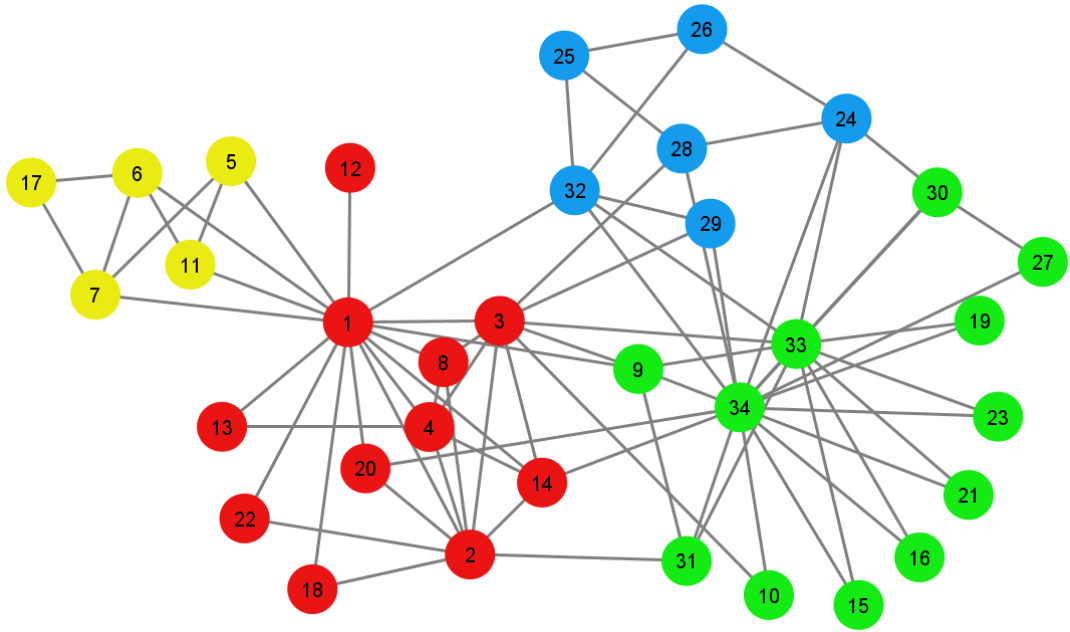

Figure 5. Result of applying the Louvain algorithm with multilevel refinement to the karate club network. Final solution in the original network.

**SmartLocalMovingAlgorithm**

**input:**

    *A*: Adjacency matrix of a network

    *c*: Initial assignment of nodes to communities

**output:**

    *c*: Final assignment of nodes to communities

*// Run the local moving heuristic.*

$c \leftarrow$ LocalMovingHeuristic($A$, $c$)

**if** NumberOfCommunities($c$) < NumberOfNodes($A$) **then**

    *// For each community, construct a subnetwork and run the local moving heuristic.*

    *// Construct a reduced network based on the community structure of the subnetworks.*

    $c_{old} \leftarrow c$

    $j \leftarrow 0$

    **for** $i \leftarrow 1$ **to** NumberOfCommunities($c_{old}$) **do**

        $A_{sub} \leftarrow$ Subnetwork($A$, $c_{old}$, $i$)

        $c_{sub} \leftarrow [1…$NumberOfNodes($A_{sub}$)$]$

        $c_{sub} \leftarrow$ LocalMovingHeuristic($A_{sub}$, $c_{sub}$)

        $c(c_{old} = i) \leftarrow c_{sub} + j$

        $c_{reduced}([j + 1]…[j +$ NumberOfCommunities($c_{sub}$)$]) \leftarrow i$

        $j \leftarrow j +$ NumberOfCommunities($c_{sub}$)

    **end for**

    $A_{reduced} \leftarrow$ ReducedNetwork($A$, $c$)

    *// Perform a recursive call to identify the community structure of the reduced network.*

    $c_{reduced} \leftarrow$ SmartLocalMovingAlgorithm($A_{reduced}$, $c_{reduced}$)

    *// Merge communities based on the community structure of the reduced network.*

    $c_{old} \leftarrow c$

    **for** $i \leftarrow 1$ **to** NumberOfCommunities($c_{old}$) **do**

        $c(c_{old} = i) \leftarrow c_{reduced}(i)$

    **end for**

**end if**

Figure 6. Summary of the main steps of the SLM algorithm.

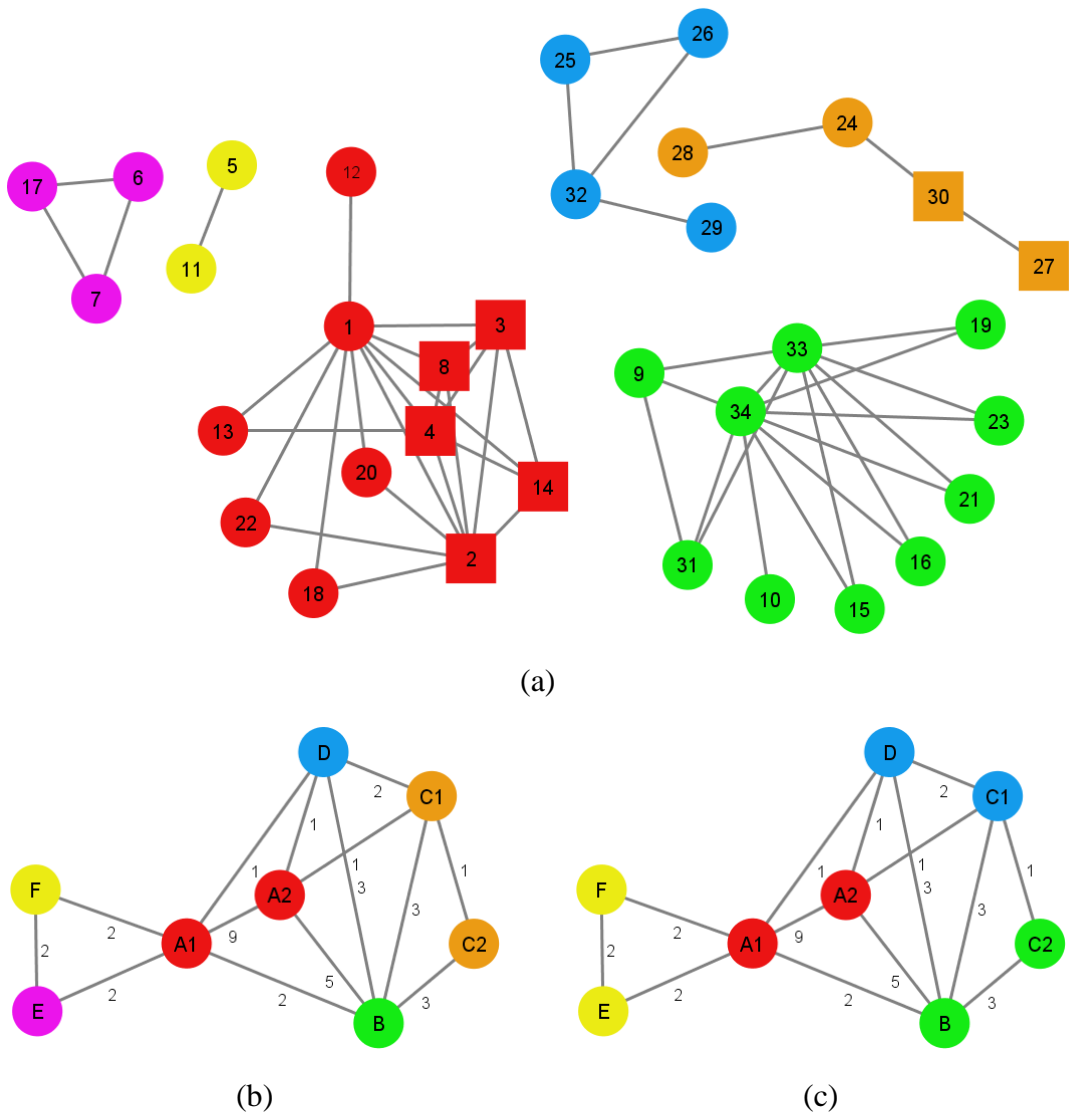

(a)

(b)                                    (c)

Figure 7. Result of applying the SLM algorithm to the karate club network. (a) Six subnetworks. Using the local moving heuristic, the red and orange subnetworks have been split up into two communities. Nodes in these subnetworks are displayed using either a circle or a square, depending on the community to which they belong. (b) Reduced network before applying the local moving heuristic. (c) Reduced network after applying the local moving heuristic. Notice that self links in the reduced network are not shown.

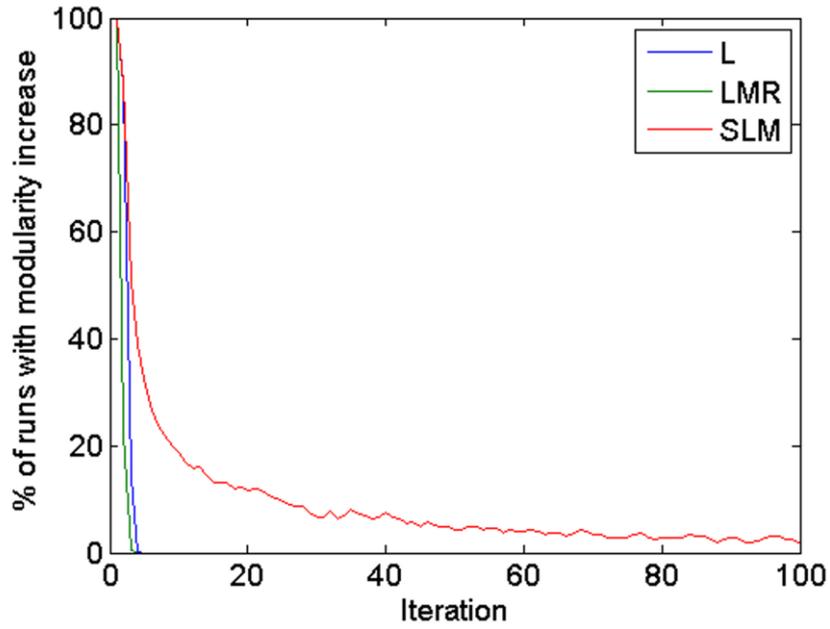

Figure 8. Effect of performing multiple iterations of an algorithm for 13 small and medium-sized networks. For each iteration, the percentage of all 1300 algorithm runs that resulted in a modularity increase is shown. The algorithms are the original Louvain algorithm (L), the Louvain algorithm with multilevel refinement (LMR), and the SLM algorithm.

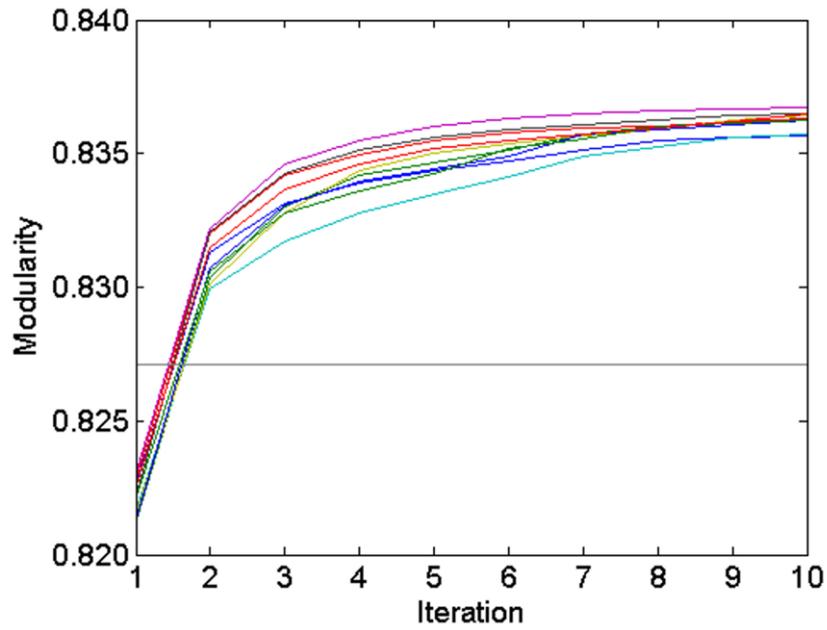

(a)

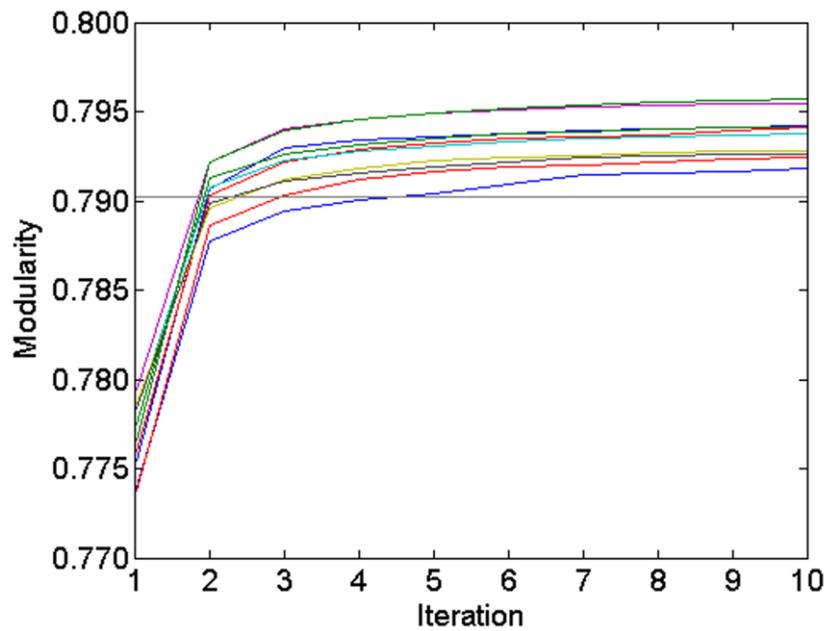

(b)

Figure 9. Modularity value at the end of each of the 10 iterations in each of the 10 runs of the SLM algorithm. The horizontal line indicates the highest modularity value obtained using the iterative variants of the original Louvain algorithm and the Louvain algorithm with multilevel refinement. (a) DBLP network. (b) WoS network.

Table 1. Results for 13 small and medium-sized networks. For each network, the number of nodes and edges is reported as well as the highest modularity value $Q_{CSA}$ obtained using the CSA algorithm of Lee et al. (2012). In addition, results are reported for the original Louvain algorithm (L), the Louvain algorithm with multilevel refinement (LMR), and the SLM algorithm. These results are based on 100 algorithm runs consisting of either one or 100 iterations. The values that are shown are the differences between the highest modularity values obtained using the L, LMR, and SLM algorithms and the highest modularity values obtained using the CSA algorithm. Negative (positive) values indicate that an algorithm is performing worse (better) than the CSA algorithm. In the case of an equal performance, no value is shown.

| | Nodes | Edges | $Q_{CSA}$ | 1 iteration | | | 100 iterations | | |
| --- | --- | --- | --- | --- | --- | --- | --- | --- | --- |
| | | | | L | LMR | SLM | L | LMR | SLM |
| Dolphins | 62 | 159 | 0.5285 | | | | | | |
| Les Misérables | 77 | 254 | 0.5600 | | | | | | |
| Political books | 105 | 441 | 0.5272 | | | | | | |
| College football | 115 | 613 | 0.6046 | | | | | | |
| Jazz | 198 | 2 742 | 0.4451 | | | | | | |
| USAir97 | 332 | 2 126 | 0.3682 | -0.0015 | -0.0006 | -0.0002 | -0.0006 | -0.0006 | |
| Netscience_main | 379 | 914 | 0.8486 | -0.0004 | -0.0002 | -0.0005 | -0.0002 | -0.0002 | |
| C. elegans | 453 | 2 025 | 0.4533 | -0.0060 | -0.0045 | -0.0064 | -0.0045 | -0.0045 | -0.0004 |
| Electronic circuit (s838) | 512 | 819 | 0.8194 | -0.0157 | -0.0039 | -0.0197 | -0.0039 | -0.0039 | -0.0018 |
| E-mail | 1 133 | 5 451 | 0.5828 | -0.0055 | -0.0021 | -0.0061 | -0.0021 | -0.0021 | |
| Erdos02 | 6 927 | 11 850 | 0.7184 | -0.0189 | -0.0041 | -0.0182 | -0.0041 | -0.0041 | -0.0005 |
| PGP | 10 680 | 24 316 | 0.8867 | -0.0027 | -0.0018 | -0.0021 | -0.0018 | -0.0018 | -0.0004 |
| condmat2003 | 27 519 | 116 181 | 0.7675 | -0.0088 | -0.0041 | -0.0083 | -0.0041 | -0.0041 | 0.0039 |

Table 2. Results for six large networks. For each network, the number of nodes and edges is shown in the first column. Results are reported for the original Louvain algorithm (L), the Louvain algorithm with multilevel refinement (LMR), and the SLM algorithm. These results are based on 10 algorithm runs consisting of either one or 10 iterations. $Q_{min}$ and $Q_{max}$ denote, respectively, the lowest and the highest modularity value obtained in 10 algorithm runs, and $t$ denotes the average computing time per algorithm run (in seconds).

| | | 1 iteration | | | 10 iterations | | |
|---|---|---|---|---|---|---|---|
| | | L | LMR | SLM | L | LMR | SLM |
| Amazon (0.5M / 0.9M) | $Q_{min}$ | 0.9257 | 0.9293 | 0.9261 | 0.9293 | 0.9293 | 0.9335 |
| | $Q_{max}$ | 0.9264 | 0.9298 | 0.9267 | 0.9299 | 0.9299 | 0.9338 |
| | $t$ | 6 | 7 | 7 | 9 | 9 | 28 |
| DBLP (0.4M / 1.0M) | $Q_{min}$ | 0.8203 | 0.8243 | 0.8213 | 0.8243 | 0.8243 | 0.8357 |
| | $Q_{max}$ | 0.8227 | 0.8271 | 0.8231 | 0.8271 | 0.8271 | 0.8367 |
| | $t$ | 7 | 8 | 8 | 9 | 9 | 26 |
| IMDb (0.4M / 15.0M) | $Q_{min}$ | 0.6976 | 0.6994 | 0.6978 | 0.6994 | 0.6994 | 0.7050 |
| | $Q_{max}$ | 0.7041 | 0.7051 | 0.7061 | 0.7052 | 0.7052 | 0.7077 |
| | $t$ | 18 | 22 | 23 | 26 | 26 | 100 |
| LiveJournal (4.0M / 34.7M) | $Q_{min}$ | 0.7441 | 0.7576 | 0.7473 | 0.7578 | 0.7578 | 0.7676 |
| | $Q_{max}$ | 0.7557 | 0.7658 | 0.7568 | 0.7658 | 0.7658 | 0.7720 |
| | $t$ | 350 | 505 | 375 | 566 | 582 | 1 549 |
| WoS (10.6M / 104.5M) | $Q_{min}$ | 0.7714 | 0.7851 | 0.7736 | 0.7851 | 0.7851 | 0.7918 |
| | $Q_{max}$ | 0.7786 | 0.7902 | 0.7793 | 0.7902 | 0.7902 | 0.7957 |
| | $t$ | 6 800 | 8 113 | 6 812 | 8 398 | 8 415 | 19 994 |
| Web uk-2005 (39.5M / 783.0M) | $Q_{min}$ | 0.9793 | 0.9796 | 0.9793 | 0.9796 | 0.9796 | 0.9801 |
| | $Q_{max}$ | 0.9795 | 0.9797 | 0.9795 | 0.9797 | 0.9797 | 0.9801 |
| | $t$ | 11 006 | 9 698 | 10 576 | 11 736 | 9 993 | 17 074 |